\documentclass{webofc}
\usepackage[varg]{txfonts}   
\usepackage{xcolor}

\newcommand{\snn}   {\sqrt{s_{_{NN}}}}
\newcommand{\avfd}  {\textsc{avfd}}
\newcommand{\ampt}  {\textsc{ampt}}
\newcommand{\epos}  {\textsc{epos\footnotesize{4}}}
\newcommand{\hydjet}{\textsc{hydjet\scriptsize{++}}}
\newcommand{\gevc}  {GeV/$c$}
\newcommand{\pt}    {p_{\rm T}}
\newcommand{\Ks}    {K_S}
\newcommand{\dg}    {\Delta\gamma}
\newcommand{\gos}   {\dg_{\rm os}}
\newcommand{\gss}   {\dg_{\rm ss}}
\newcommand{\cme}   {\textsc{cme}}
\newcommand{\bkg}   {\text{bkg}}
\newcommand{\res}   {\text{res}}
\newcommand{\vres}  {v_{2,{\text{res}}}}
\newcommand{\dgess} {\dg_{\textsc{ess}}}
\newcommand{\dgese} {\dg_{\textsc{ese}}}

\newcommand{\qpair} {q_{2,{\rm pair}}}
\newcommand{\vsing} {v_{2,{\rm single}}}

\newcommand\mean[1]{\left\langle#1\right\rangle}

\begin{document}
\title{Investigating the Event-Shape Methods in Search for the Chiral Magnetic Effect in Relativistic heavy-ion Collisions}
%
%

\author{\firstname{Han-Sheng} \lastname{Li}\inst{1}\fnsep\thanks{\email{li3924@purdue.edu}} \and
        \firstname{Yicheng} \lastname{Feng}\inst{1} \and
        \firstname{Fuqiang} \lastname{Wang}\inst{1}}

\institute{Department of Physics and Astronomy, Purdue University, West Lafayette, IN 47907, USA
          }

\abstract{%
    The azimuthal correlator $\dg$ searching for the chiral magnetic effect (CME) is contaminated by a major background proportional to the elliptic flow $v_2$. 
    Event-shape engineering (ESE) and event-shape selection (ESS) binning events in {\em dynamical} and {\em statistical} fluctuations of $v_2$, respectively, are two methods searching for the CME. We conduct a systematic study using physics and toy model simulations. It is found that ESE fulfills the general premise of measuring the CME but is statistically hungry, whereas ESS is not practical to measure the CME because of the intertwining variables used in the method.
}
\maketitle
\section{Introduction}\label{intro}

Interactions of quarks with topological gluon fields in quantum chromodynamics (QCD) can cause chirality imbalance in local metastable domains, breaking the parity and charge-parity symmetries. Such chirality imbalance can yield charge separation along a strong magnetic field, a phenomenon called the chiral magnetic effect (CME)~\cite{Kharzeev:1998kz}.
Ultra-strong magnetic fields are produced in non-central relativistic heavy-ion collisions~\cite{Skokov:2009qp}, making these collisions an ideal place to search for the CME~\cite{Kharzeev:2015znc,Zhao:2019hta,Feng:2025yte}. 

Because the magnetic field in heavy-ion collisions is on average perpendicular to the reaction plane (RP), a CME signal may be conveniently quantified by the $a_1$ parameter in Fourier expansion of particle azimuthal distribution~\cite{Voloshin:1994mz,Kharzeev:2015znc},
$dN_\pm/d\phi \propto 1 \pm 2a_1\sin(\phi-\psi) + 2v_2\cos2(\phi-\psi) + \cdots$,
where $\phi$ is particle's azimuthal angle, $\psi$ is that of the RP's, and the subscript `$\pm$' indicates particle charge sign. 
The elliptic (flow) anisotropy $v_2$ harmonic is the leading modulation. 
Because of the vanishing mean $\mean{a_1}$ due to random fluctuations of the chirality sign, a commonly used observable is  the three-point correlator~\cite{Voloshin:2004vk},
$\gamma=\mean{\cos(\phi_\alpha+\phi_\beta-2\psi)}$,
where $\phi_\alpha$ and $\phi_\beta$ are the azimuthal angles of two particles of interest (POIs). To cancel charge-independent backgrounds, such as effects from global momentum conservation, the difference between opposite-sign (OS) and same-sign (SS) correlators is used~\cite{Abelev:2009ac}, $\dg \equiv \gos - \gss$. 
The CME signal presented in the $\dg$ observable would then be $\dg_\cme=\mean{2a_1^2}$.

However, charge-dependent correlations also exist, including resonance decays and jet correlations, which contribute more to OS than SS pairs. 
These correlations are turned into RP-dependent background by elliptic flow of heavy-ion collisions. 
Such $v_2$-induced backgrounds can generally be expressed by
$\dg_\bkg\propto\mean{\cos(\phi_\alpha+\phi_\beta-2\phi_\res)}v_{2,\res}$,
where $\phi_\res$ and $v_{2,\res}$ are the azimuthal angle and elliptic flow of the background source. 
This property is exploited by two analysis methods, event-shape engineering (ESE) and event-shape selection (ESE), in attempt to remove the $v_2$-induced backgrounds.

\section{Methodologies}\label{method}

\begin{figure*}[hbt]
\begin{minipage}{0.75\textwidth}
    \includegraphics[width=0.5\textwidth]{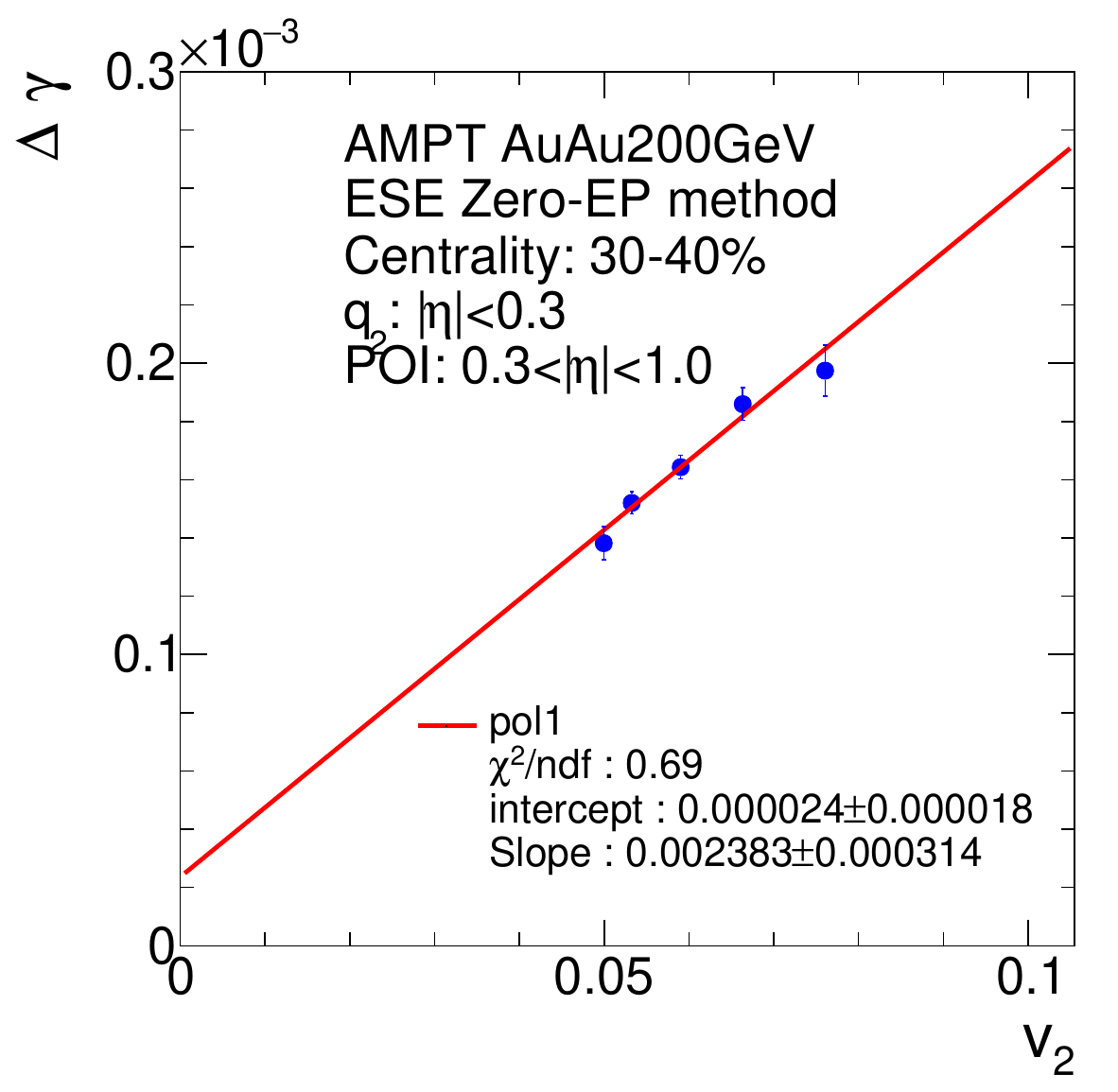}\hfill
    \includegraphics[width=0.5\textwidth]{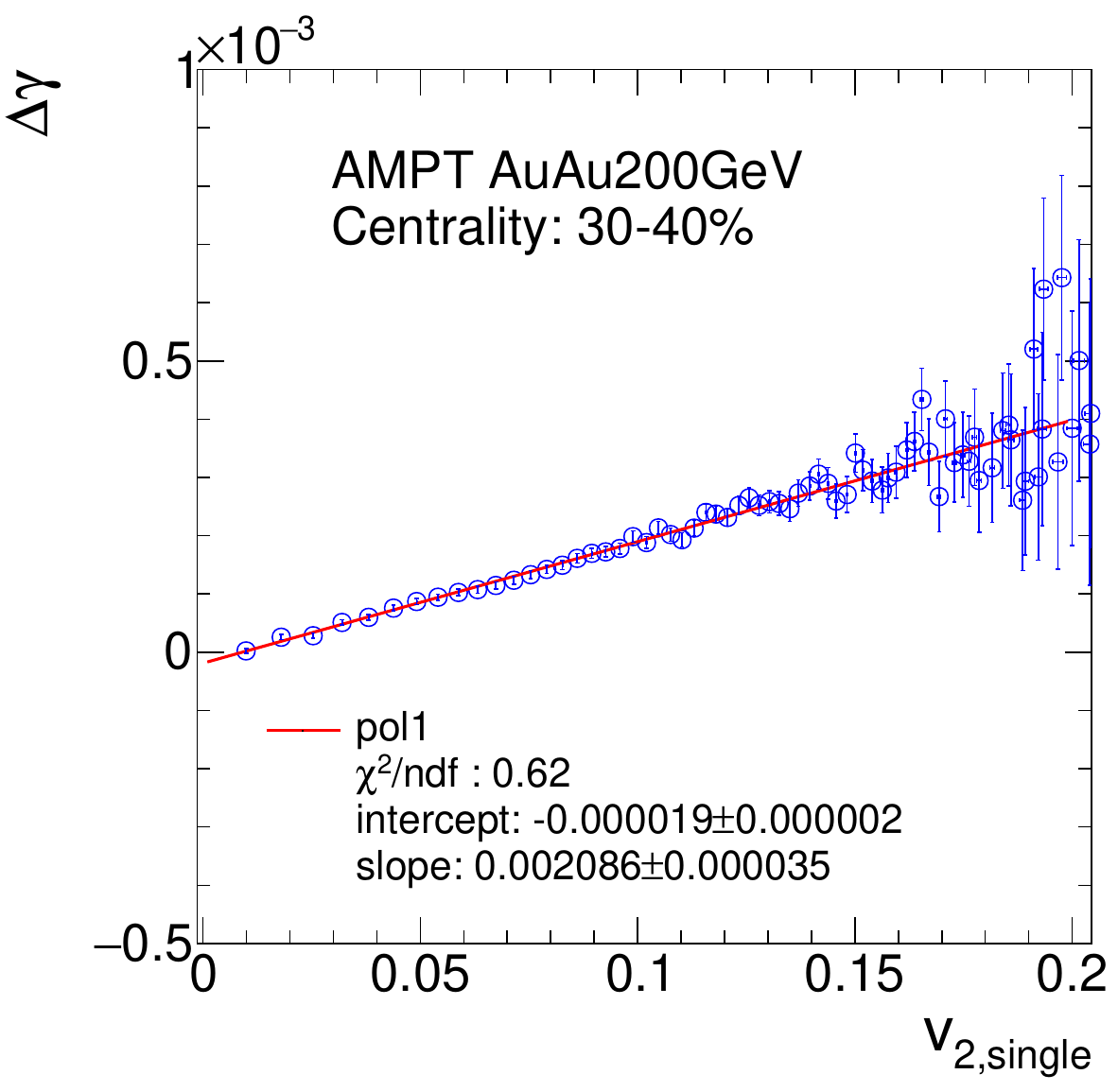}\hfill
\end{minipage}\hfill
\begin{minipage}{0.23\textwidth}
    \caption{Example results of $\dg$ versus $v_2$ and linear fits by Eq.~(\ref{eq:fit}) from the ESE (left) and ESS (right) methods. Example is taken from \ampt\ simulation of 30--40\% centrality Au+Au collisions at $\snn=200$~GeV.}
    \label{fig:fits}
\end{minipage}
\end{figure*}

The general idea of the ESE and ESS methods is the same: collision events with similar CME signals (within a given narrow centrality bin) are grouped according to varying $v_2$ by specific means/quantity and the $\dg$ observable is measured in individual groups,
\begin{equation}
    \dg(v_2) = \dg_\cme + k v_2 \,.
    \label{eq:fit}
\end{equation}
The intercept of a linear fit to $\dg(v_2)$ would yield the CME signal $\dg_\cme$.
Examples are shown in Fig.~\ref{fig:fits}; we use $\dgese$ and $\dgess$ to denote the corresponding intercepts.

The two methods differ in the quantity used to group events.
Both calculate the event-by-event flow vector, 
$\vec{q}_2 = \frac{1}{\sqrt{N}} \left( \sum_{i=1}^{N} \cos2\phi_i, \sum_{i=1}^{N} \sin2\phi_i \right)$,
and use its magnitude 
$q_2 = \left[1 + \frac{1}{N}\sum_{i\neq j} \cos 2(\phi_i-\phi_j)\right]^{1/2}$ to group events.
However, ESE~\cite{Schukraft:2012ah} calculates $q_2$ using particles {\em different} from the POIs, whereas ESS calculates $\qpair$ of pairs of the {\em very} POIs~\cite{Xu:2023elq}.
%
As a result, the former selects events on {\em dynamical} fluctuations of $\mean{v_2}$ of the POIs, whereas the latter selects events primarily on {\em statistical} fluctuations of $v_2$ (whose average is biased):
\begin{itemize}
    \item Stemming from the same initial geometry, the dynamically fluctuated $\mean{v_2}$'s of CME background sources are all proportional to the final-state particle $\mean{v_2}$; thus, the projection to $\mean{v_2}=0$ is free of background contributions. 
    This is illustrated in the left cartoon of Fig.~\ref{cartoon}, where the three ellipses indicate event selections by $q_2$ and the small circles are the average $(\mean{v_2}$, $\mean{\vres})$ in each group. 
    Because dynamical fluctuations are relatively small in narrow centrality bins, large uncertainties ensue in $\dgese$ intercepts (see left panel of Fig.~\ref{fig:fits}).
\begin{figure*}[hbt]
\begin{minipage}{0.70\textwidth}
    \centering
    \includegraphics[width=0.45\textwidth]{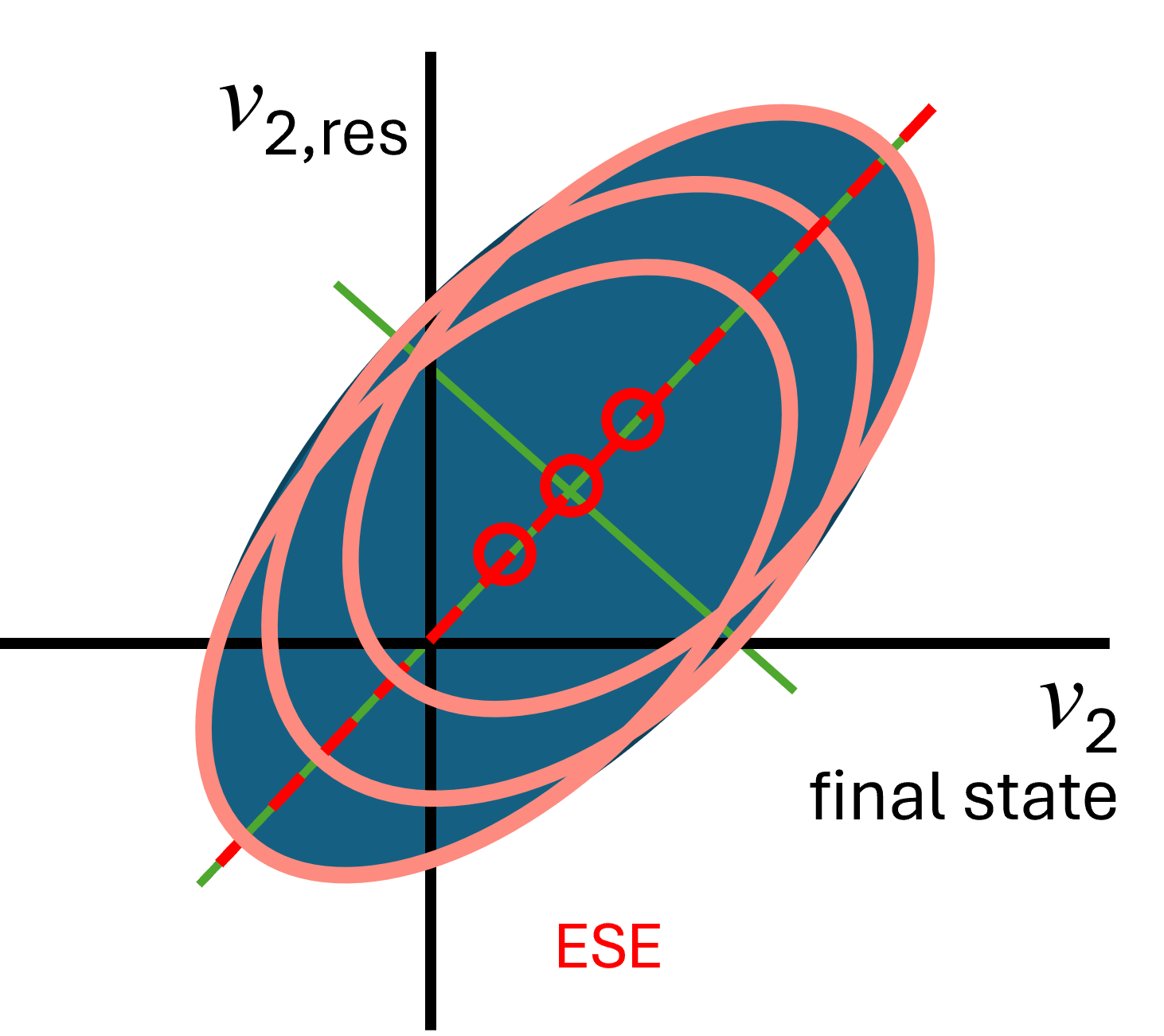}
    \includegraphics[width=0.45\textwidth]{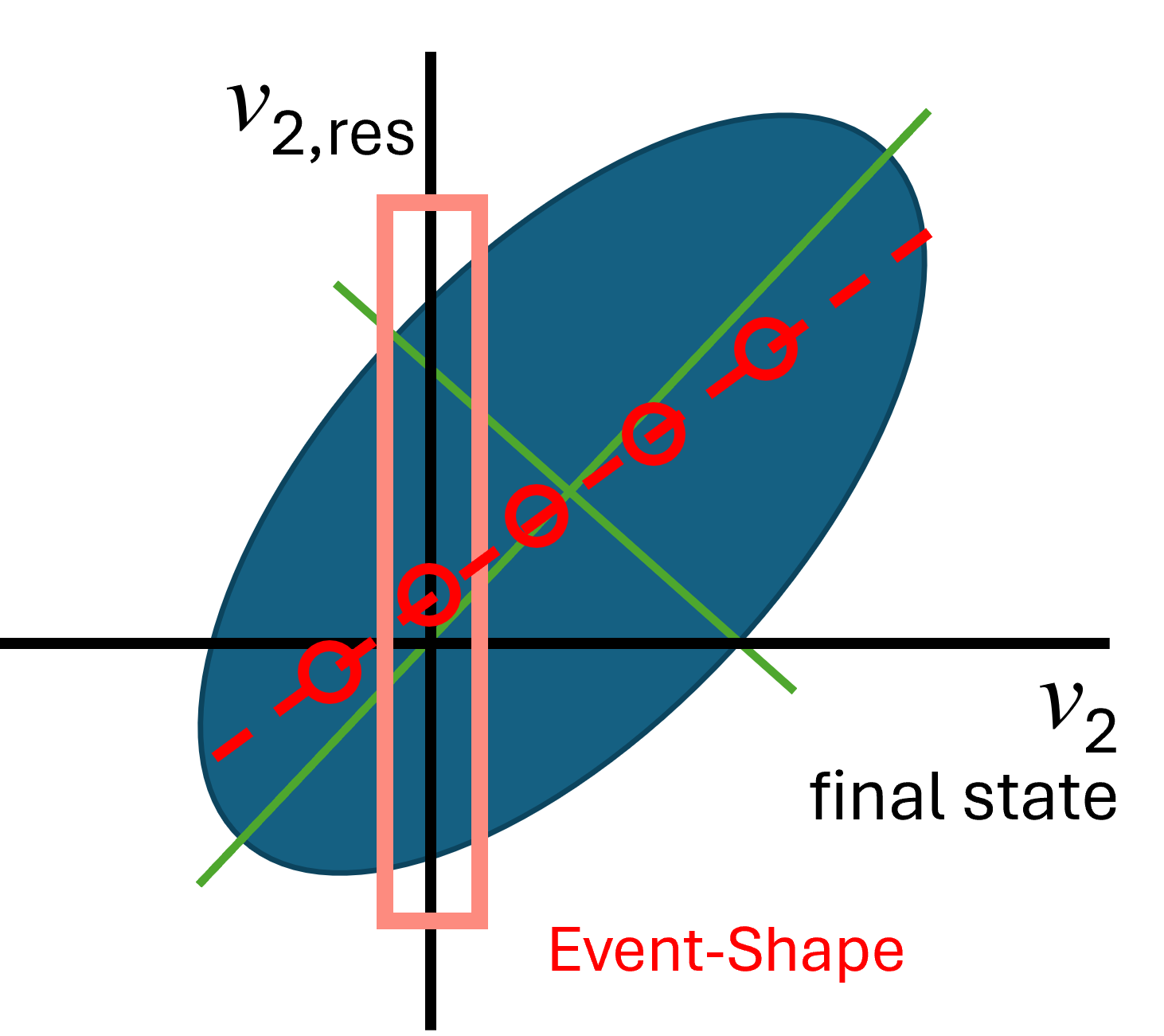}
\end{minipage}
\begin{minipage}{0.28\textwidth}
    \caption{Cartoons illustrating the dynamical (left) and statistical (right) event-shape methods in terms of the event-by-event resonance $\vres$ versus final-state single particle $v_2$ in events within a given narrow centrality bin.} 
    \label{cartoon}
\end{minipage}
\end{figure*}
    \item 
    Because of the statistical fluctuation nature (as clear from the wide range of $\vsing$ in Fig.~\ref{fig:fits} right panel), the mean $\vres$'s of background sources in the ESS method are not guaranteed to be proportional to the mean $\vsing$ in each event group; non-zero backgrounds can remain at $\vsing=0$.
    A simple case of ESS is using the $v_2$ itself as the event-selection quantity, published by STAR in 2014~\cite{Adamczyk:2013kcb}, as illustrated by the right cartoon of Fig.~\ref{cartoon}. 
\end{itemize}

\section{Results and Discussion}
We have conducted a systematic study using four heavy-ion physics models, \avfd, \ampt, \epos, and \hydjet, simulating Au+Au collisions at nucleon-nucleon center-of-mass energy $\snn=200$ and 27~GeV. 
The full description of the study is documented in Li et al.~\cite{Li:2024gdz}.

\begin{figure*}[hbt]
\begin{minipage}{0.75\textwidth}
    \includegraphics[width=0.5\textwidth]{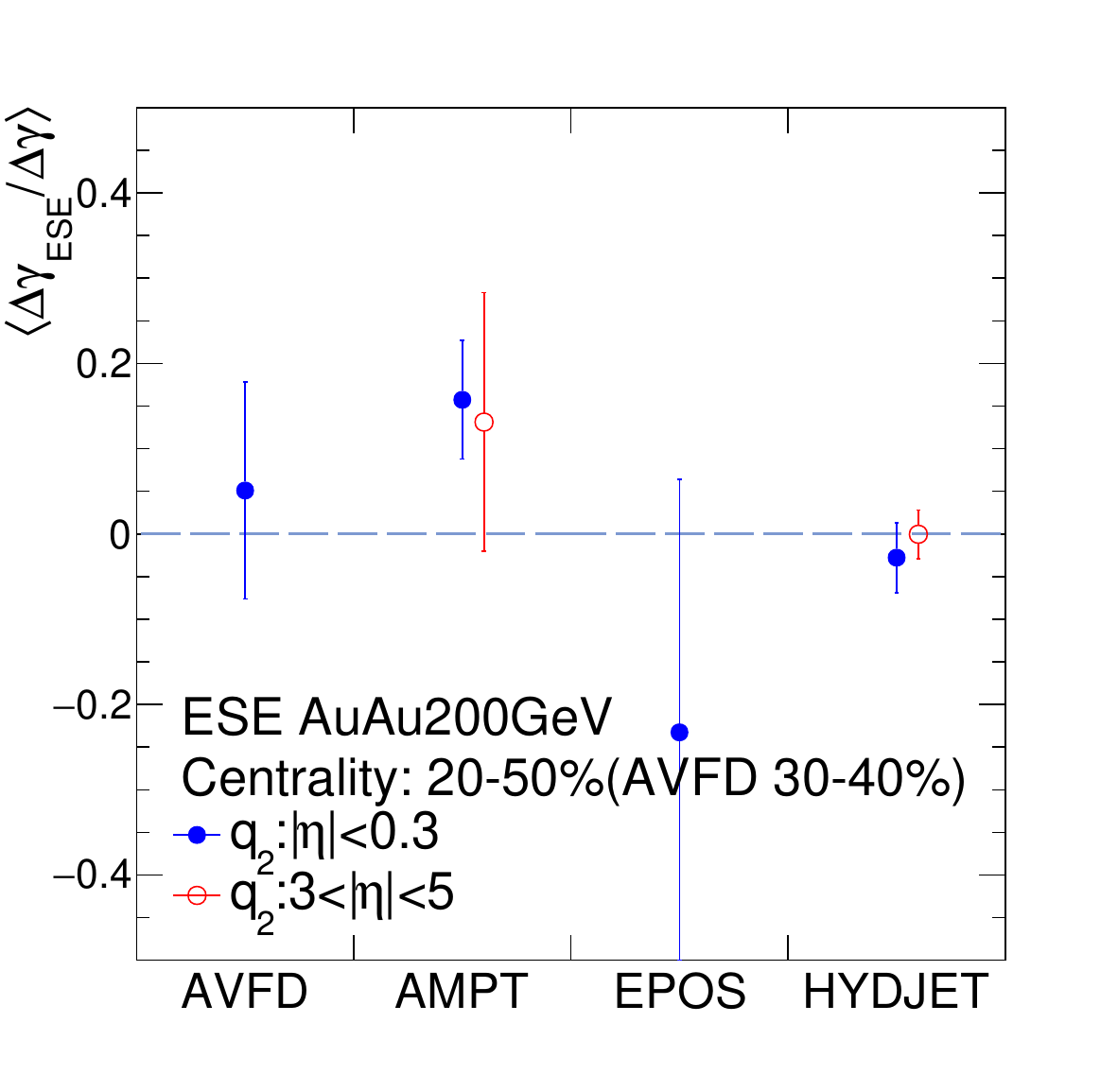}\hfill
    \includegraphics[width=0.5\textwidth]{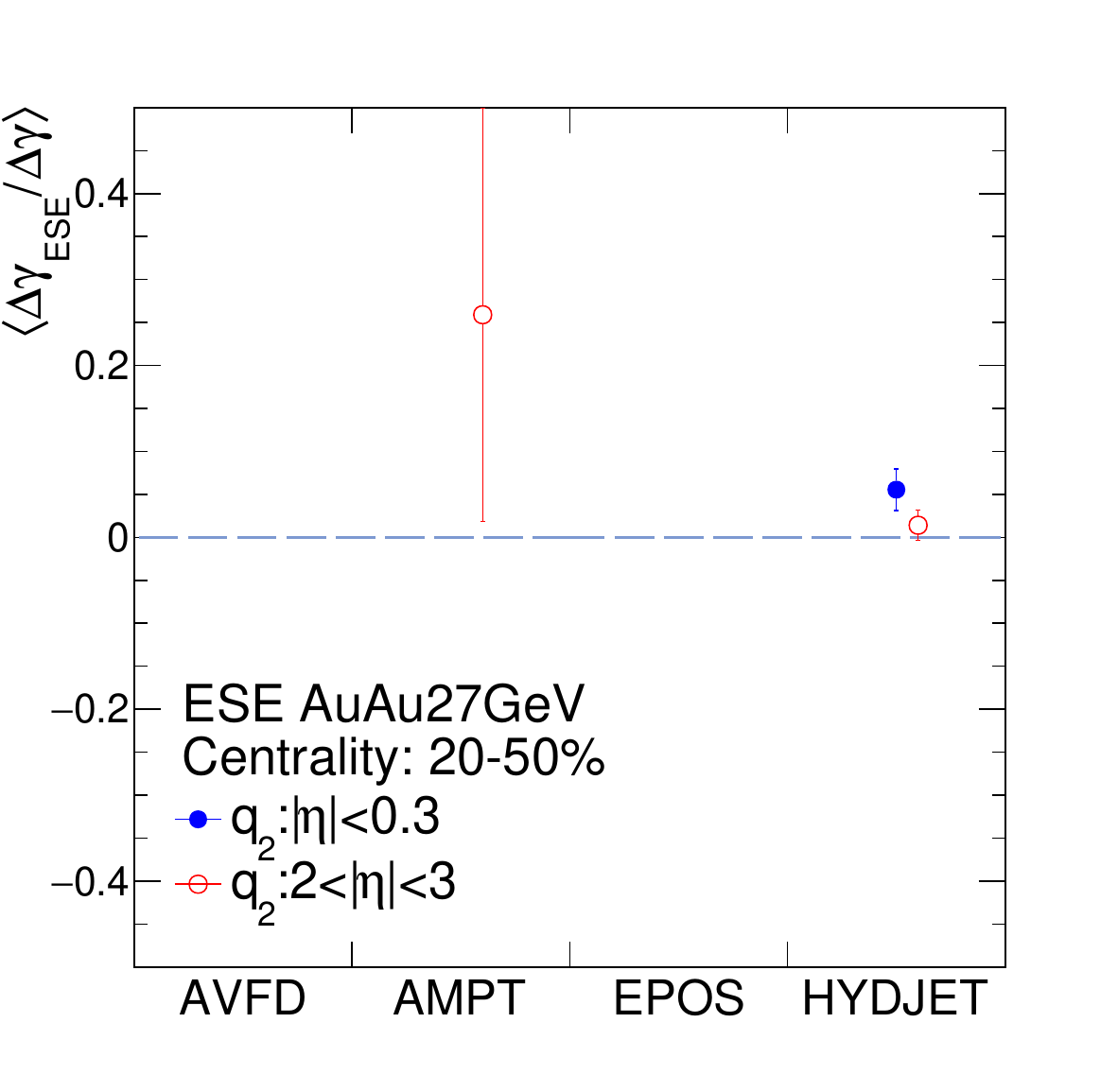}\hfill
    \includegraphics[width=0.5\textwidth]{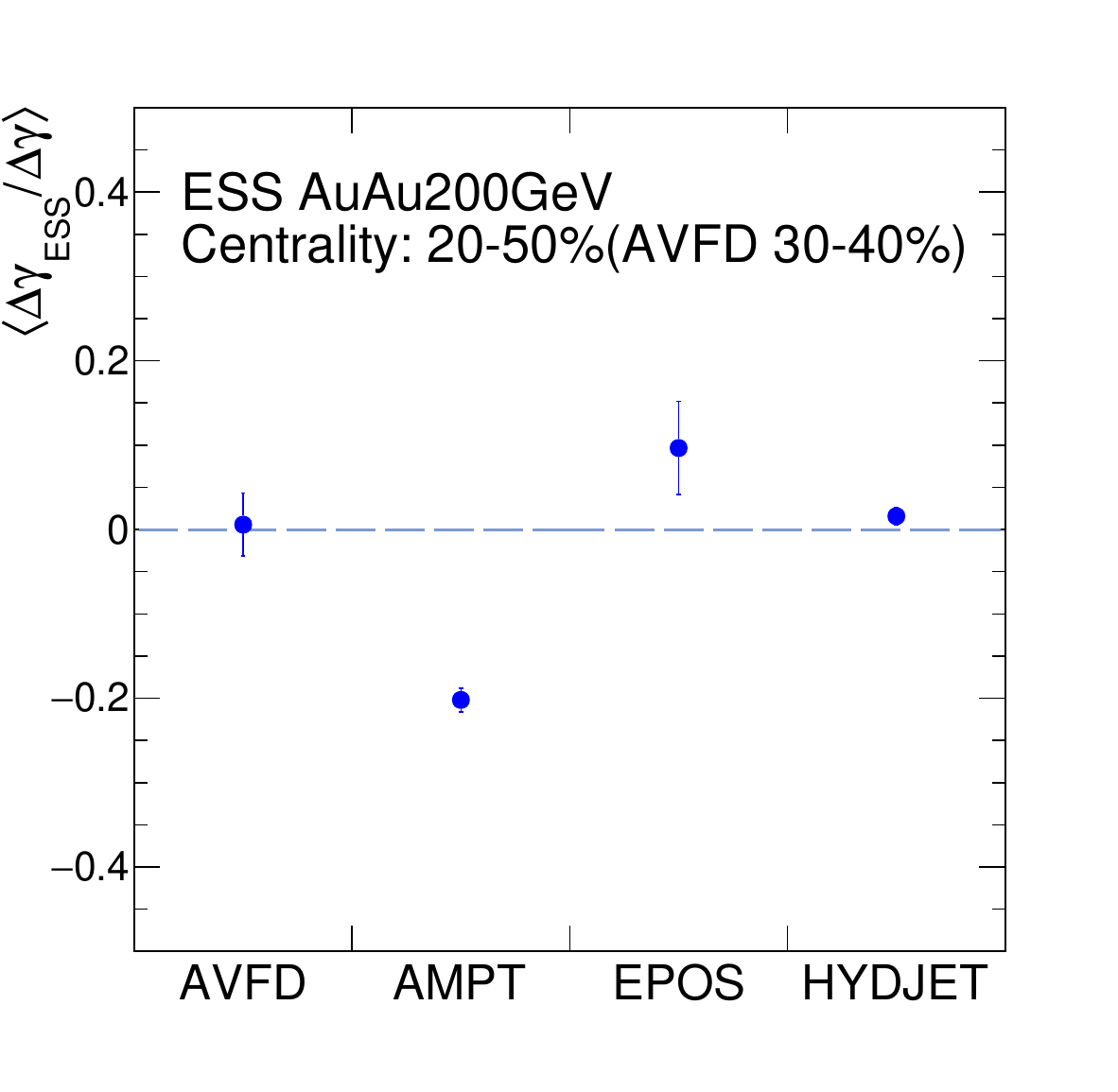}\hfill
    \includegraphics[width=0.5\textwidth]{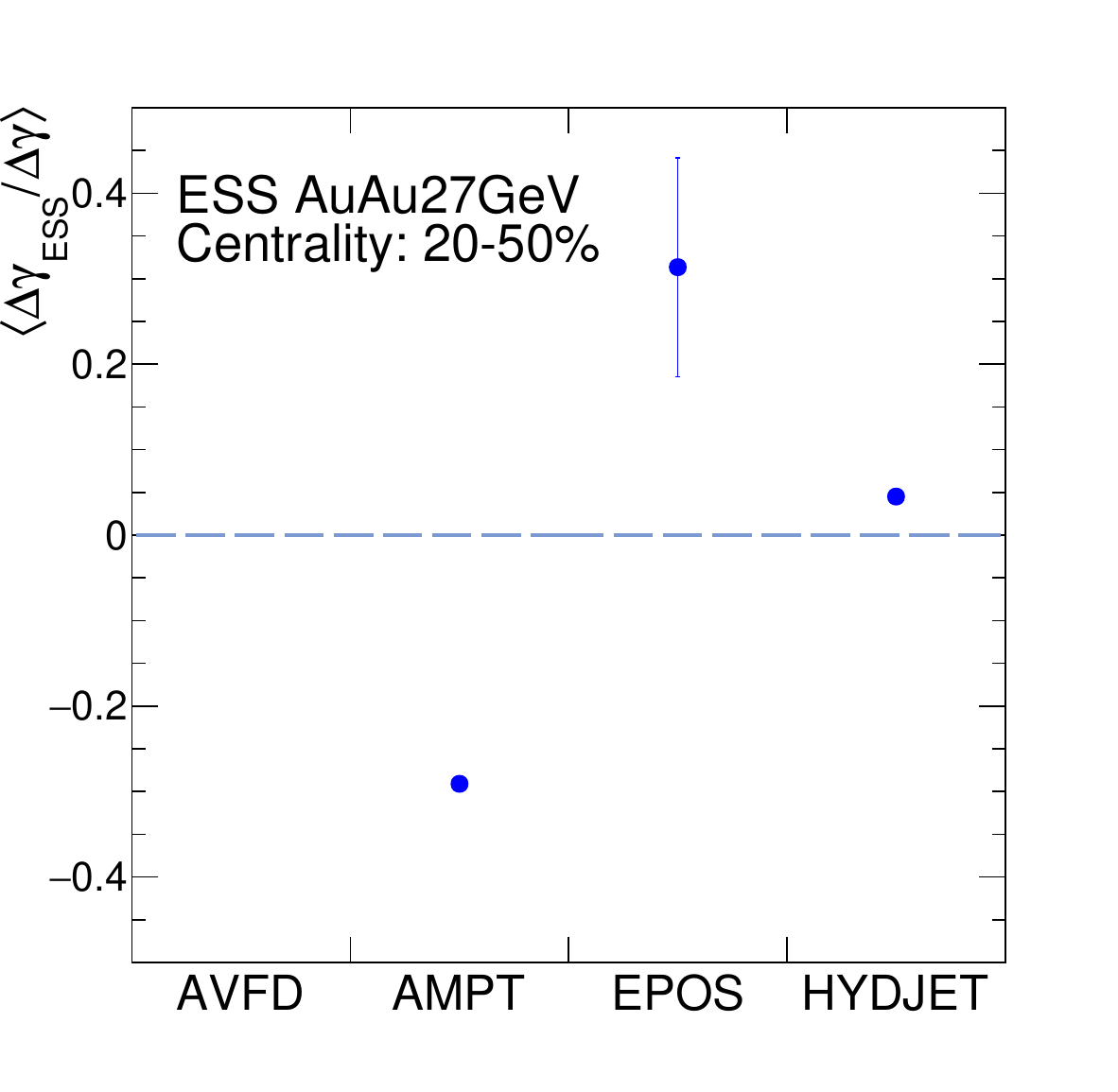}\hfill
    \label{fig:inter}
\end{minipage}
\hfill
\begin{minipage}{0.24\textwidth}
    \caption{ESE (upper panels) and ESS (lower panels) intercepts divided by the corresponding inclusive $\dg$ values in Au+Au collisions at $\snn=200$~GeV (left panels) and 27~GeV (right panels) simulated by four physics models. ESE uses POIs from $0.3<|\eta|<1$ and $q_2$ calculated by particles in $|\eta|<0.3$ or a forward/backard region (for \ampt\ and \hydjet). ESS uses POIs from $|\eta|<1$ ($|\eta|<2$ for \epos) which are also used for $\qpair$ calculation. The $\pt$ range is $0.2<\pt<2$~\gevc\ for all studies.}
\end{minipage}
\end{figure*}

Figure~\ref{fig:inter} summarizes the intercepts divided by the overall $\dg$ magnitude from ESE and ESS. 
No CME signal is present in any of the model simulations shown in these proceedings, so ideally all intercepts should be zero if the methods worked as promised.

ESE uses POIs from $0.3<|\eta|<1$ and particles from a different momentum space for $q_2$ calculation. The $\dgese$ intercepts are mostly consistent with zero, suggesting that the ESE method is doing a proper job.
With $q_2$ from $|\eta|<0.3$, the $\dgese$ values are $\sim2$ standard deviations above zero for \ampt\ at 200~GeV and \hydjet\ at 27~GeV; using $q_2$ from forward/backward $\eta$ regions yield $\dgese$ consistent with zero. We thus postulate the cause for the $2\sigma$ deviations to be nonflow correlations between POIs and $q_2$ not far enough apart in $\eta$. 

ESS uses POIs from midrapidity and calculates $\qpair$ of pairs of these POIs.
The $\dgess$ intercepts are not always consistent with zero, as shown in the lower panels of Fig.~\ref{fig:inter}. It appears that the $\dgess$ value can be positive, zero, or negative, depending on individual models. This suggests that the ESS method is {\em not} doing a proper job. 

It is noteworthy that the statistical uncertainties on $\dgese$ are larger than on $\dgess$ because of reasons mentioned in Sect.~\ref{method}. The $\dgese$ is consistent with zero but also consistent with an intercept as large as those seen by the ESS method. This is an issue of precision and can be improved with accumulation of statistics. The $\dgess$ is {\em inconsistent} with zero with the achieved statistical precision (owing to statistical fluctuations embracing $\vsing=0$). This is a demonstration of invalidity of the ESS method in obtaining background-free CME signal.

\begin{figure*}[hbt]
\begin{minipage}{0.75\textwidth}
    \includegraphics[width=0.5\textwidth]{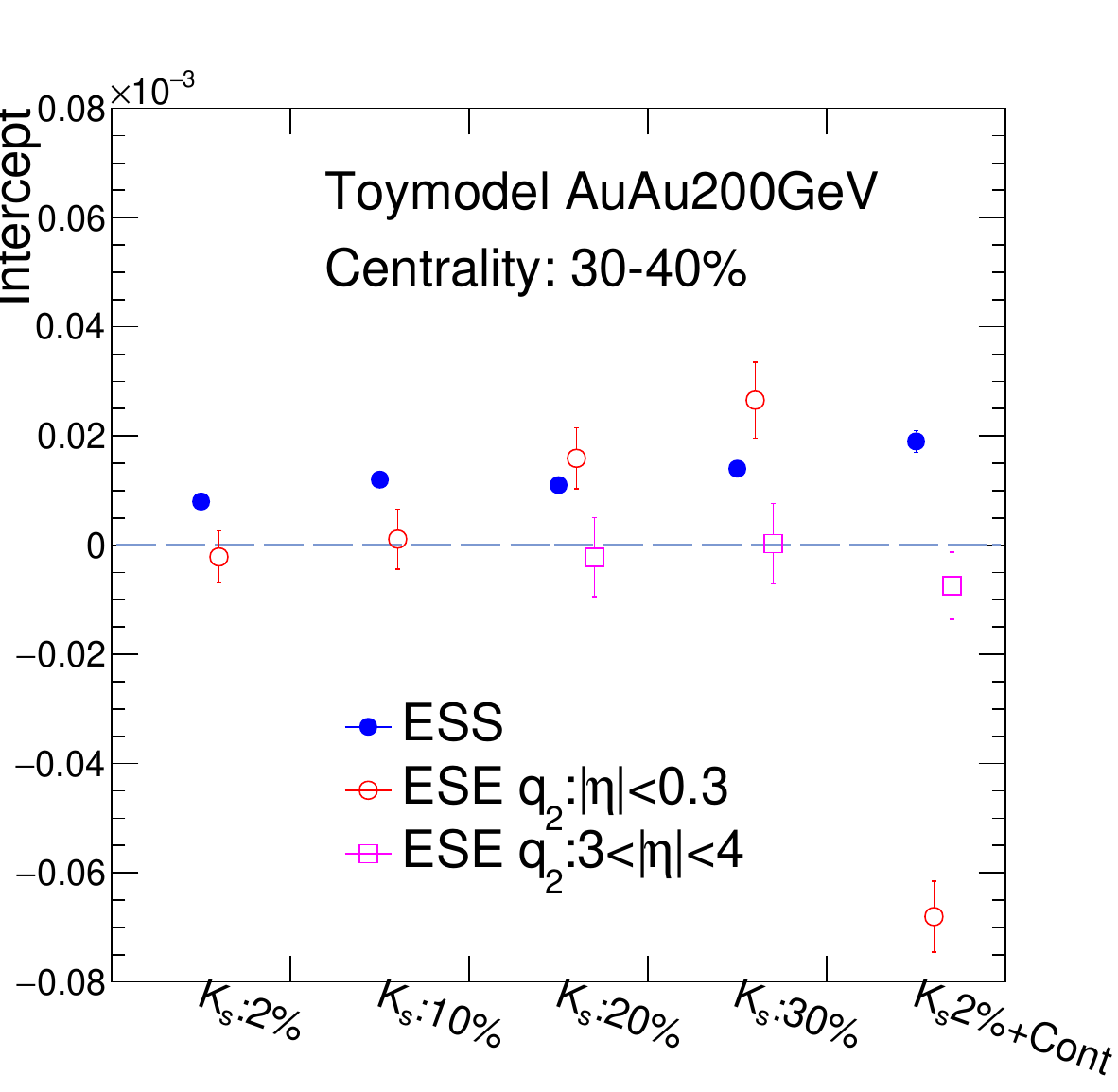}\hfill
    \includegraphics[width=0.5\textwidth]{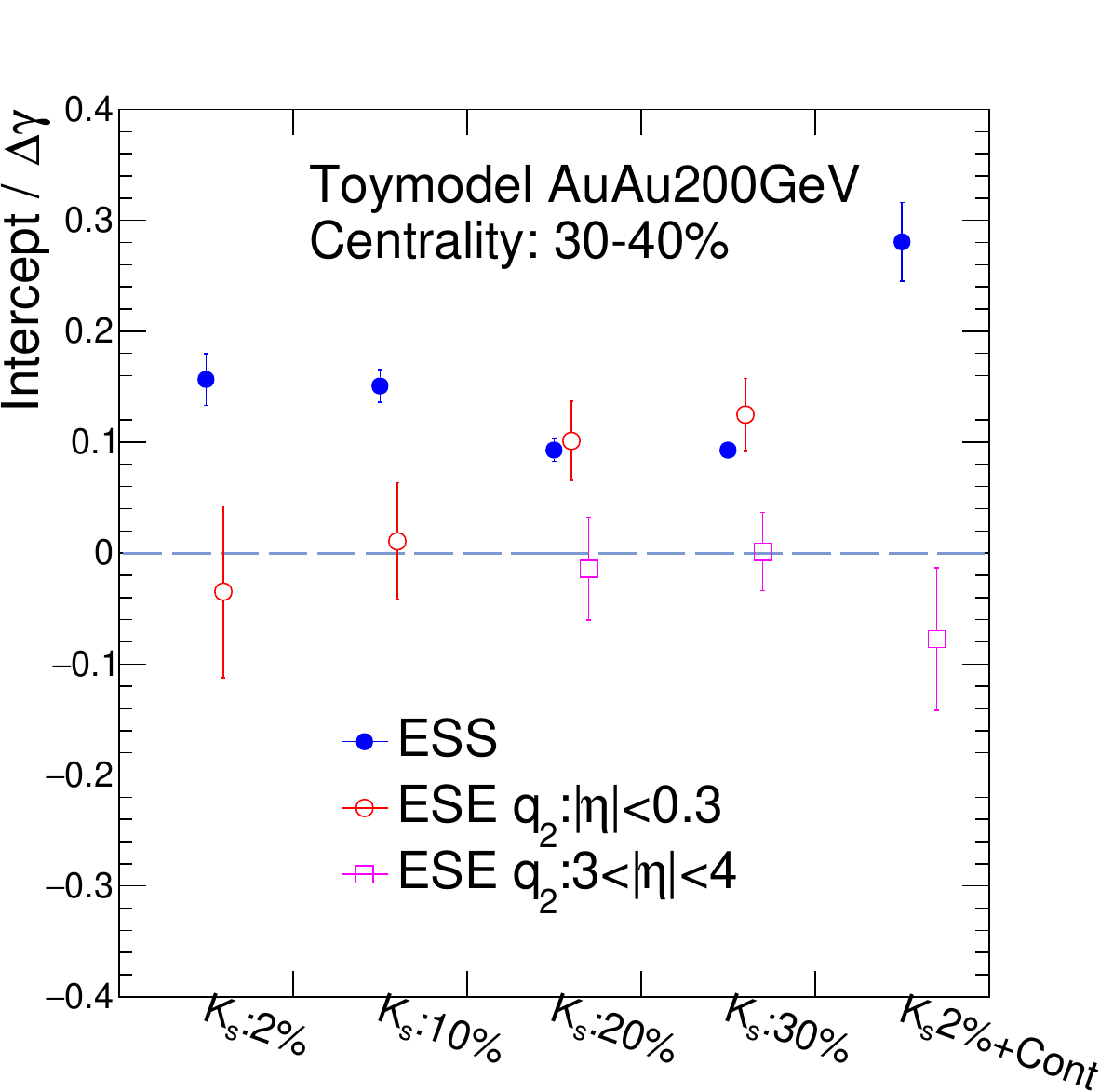}
\end{minipage}\hfill
\begin{minipage}{0.24\textwidth}
    \caption{Intercepts (left) and intercepts divided by the inclusive $\dg$ value (right) from the ESE and ESS methods as a function of the input $\Ks$ contamination in toy model simulations of 30--40\% centrality Au+Au events at $\snn=200$~GeV.}
    \label{fig:Ks}
\end{minipage}
\end{figure*}

To possibly understand further, we utilize a toy model including various $\Ks$ contributions realizing it is an important difference among the models~\cite{Li:2024gdz}. 
Figure~\ref{fig:Ks} summarizes the intercepts from the toy model study. 
The ESE intercepts are consistent with zero as long as the nonflow effects are mitigated between POIs and $q_2$. 
The ESS intercepts are all positive with varying magnitudes, suggesting that $\dgess$ depends on the event details. 
The failure of the ESS method is presumably because of correlations between $\vsing$ and $\qpair$. These correlations are inherited in the ESS method, cannot be removed, and include both flow and nonflow correlations because $\vsing$ and $\qpair$ are intertwined by using the same POIs.

\section{Summary}
The ESE and ESS methods have been exploited to search for the CME by projecting $\dg$ to $v_2=0$, relying on {\em dynamical} and {\em statistical} fluctuations of $v_2$, respectively. We have conducted a systematic study using physics models and toy models. 
It is found, with no input CME signal, that the ESE intercept is mostly consistent with zero, as expected, albeit inherently large statistical uncertainties.
The ESS intercept, on the other hand, can be positive, zero, or negative depending on the details of the simulated events. 
This suggests that one does not quantitatively know what relative contributions of CME signal and physics background are contained in the ESS measurement, because of the intertwining variables used in ESS. It is therefore not practically useful to use the ESS method to search for the CME.

\vspace{0.1in}
\noindent{\em Acknowledgment.} 
This work is supported in part by the U.S.~Department of Energy (Grant No.~DE-SC0012910).

%
%
%
\bibliography{ref}
\end{document}